\documentclass[12pt,titlepage]{article} 
 
\usepackage{setspace}  
\usepackage{geometry} 
\usepackage{amsmath} 
 
\usepackage{graphicx,color} 
 
\usepackage[round,numbers,sort&compress]{natbib}  
\usepackage[version=3]{mhchem} 
 
\newcommand{\degC}[0]{$^\circ$C}

\title{A minimal titration modelization of the
mammalian dynamical heat shock response}
 
\author{A. Siv\'ery, E. Courtade, Q. Thommen} 

\date{\today} 
 
\begin{document} 
 
\maketitle 
 
\abstract{ 
Environmental stress, such as oxidative or heat stress, induces the activation of the Heat Shock Response (HSR) which leads to an increase in the heat shock proteins (HSPs) level. These HSPs act as molecular chaperones to maintain proteostasis. Even if the main heat shock response partners are well known, a detailed description of the dynamical properties of the HSR network is still missing. In this study, we derive a minimal mathematical model of cellular response to heat shock that reproduces available experimental data sets both on transcription factor activity and cell viability. This simplistic model highlights the key mechanistic processes that rule the HSR network and reveals (i) the titration of Heat Shock Factor 1 (HSF1) by chaperones as the guiding line of the network, (ii) that protein triage governs the fate of damaged proteins and (iii) three different temperature regimes describing normal, acute or chronic stress.

 
\emph{Key words:} mammalian heat shock response; mathematical modeling; chaperones; cell stress;  
heat shock proteins; signaling pathways} 
 
\clearpage


\section*{Introduction} 
 
Molecular chaperones are key proteins of the cell machinery that
ensure correct protein folding.  The ability of a protein to
accomplish a biological function depends severely on its conformation which can lead to a loss of functionality.  The
\textit{Misfolded Proteins} (MFPs) aggregate to form stable complex
that accumulates within the cell; these aggregates are toxic and
involved in aging ~\cite{morimoto_2005,morimoto_2011}.  In unstressed
conditions, chaperones assist the correct folding of newly synthesized
proteins by caging the polypeptide chain under formation.  Chaperones
are also involved in complex mechanisms that are designed to repair
the folding injury created upon stress condition by preventing the MFP
aggregation, extracting MFP from
aggregate~\cite{liberek2008chaperones}, assisting the MFP refolding in
a ATP dependent manner~\cite{bukau2006molecular}, and lastly tagging
MFP for degradation through the ubiquitin pathway when the refolding
process failed~\cite{marques2006triage}.  Owing to their multiple
functions, chaperones are a huge protein family and represent up to one
percent of the total amount of protein within a cell in standard
conditions~\cite{csermely199890}.  Research on chaperones
inhibition is an active clinical strategy in cancer
therapies~\cite{ciocca2005heat,drysdale2006targeting}, because their
constitutive overexpression in cancerous cells decreases anticancer
agent efficiency by inhibiting apoptosis~\cite{garrido2006heat}.

A temperature rise increases the expression of many chaperones called
\textit{Heat Shock Proteins} (HSP)
~\cite{mosser1988coordinate}. Heat  severely affects the protein
folding~\cite{somero1995proteins} and the activation of HSP expression
allows to repair the heat induced injury.  However, a moderate
temperature elevation is still able to induce cell death, these
effects have been widely studied in the scope of cancer therapies to
reduce tumors~\cite{sapareto1984thermal,jolly2000role}. The
analysis of the cell viability to a given thermal protocols underlines
complex behaviors such as thermal adaptation~\cite{abravaya1991heat}
and thermotolerance. A short and non lethal heat shock enhances the
survival probability to a second normally lethal heat
increase~\cite{gerner1976transient}. The cell response to
temperature increase, called \textit{Heat Shock Response} (HSR) is
tightly linked to fundamental genetic networks such as cell
cycle~\cite{kuhl2000heat,helmbrecht2008chaperones}, metabolism~\cite{westerheide_2012}, or circadian
clock~\cite{Reinke08,tamaru2011synchronization}.  For instance, a heat
stress can lead to an arrest of the cell cycle progression in certain
phases and has been used to synchronize randomly proliferating cell
populations~\cite{Zeuthen1971}.

Thanks to intensive experimental studies, the genetic network
accounting for the HSR in mammalian cells is well established.  The \textit{Heat Shock Factor 1} (HSF1) activates the HSP transcription by
binding on the \textit{Heat Shock Elements} (HSEs) in the promoter
region of the \textit{HSP}
genes~\cite{mosser1988coordinate,baler1993activation,holmberg2002multisite,cotto1997hsf1,boulon2010nucleolus}.
Meanwhile, a chaperone complex titrates HSF1 and thus regulates
the transcriptional activation of
\textit{HSP}~\cite{abravaya1992human,cotto1996activation,kline1997repression,Zou1998,Guo2001}, while there is evidence of 
an activation of \textit{HSF1} binding on HSE upon micro-injection of misfolded proteins~\cite{Zou1998}.
All these results strongly support a sequestration mechanism of HSF by HSP to be the core of the heat shock response.  
A temperature increase induces misfolded proteins that monopolize \textit{HSP} and thus free \textit{HSF1} to activate the transcription of new \textit{HSP}.    
In this picture, the activation of the heat shock response is represented as a consequence of a competition between the chaperones affinity for both \textit{HSF1} and misfolded proteins. 

However, this coarse-grained description hides complex
mechanisms concerning the activation of HSF1 including trimerization,
phosphorylation, and translocation~\cite{voellmy2004mechanisms}, where
the phosphorylation can be evenly thermally
activated~\cite{rieger2005mathematical}.  The question of whether the
temperature sensor is the complex competition or the activation of
\textit{HSF1} remains open. 
Nevertheless, the activation of the DNA binding of HSF1 on
HSE upon a micro-injection of
misfolded proteins~\cite{Zou1998} strongly supports the sequestration
mechanism.

Numbers of studies have been carried out on the modelization of the HSR
network
dynamics~\cite{rieger2005mathematical,szymanska2009mathematical,petre2011simple,sriram2012detailed,rybinski2013modelling}. All
previously developed models of HSR attempt to describe experiments
from Abravaya \textit{et al.}~\cite{abravaya1991heat} and the best
description was given very recently by Sriram~\textit{et
  al.}~\cite{sriram2012detailed} with a detailed model of 27 ordinary
differential equations.  If all these models include a sequestration
mechanism, they differ as they involve the transcription and
translation
process~\cite{rieger2005mathematical,szymanska2009mathematical,rybinski2013modelling};
the HSF1 phosphorylation
process~\cite{rieger2005mathematical,sriram2012detailed}; the HSF1
trimerization
process~\cite{rieger2005mathematical,szymanska2009mathematical,petre2011simple,sriram2012detailed,rybinski2013modelling};
or the heat activated HSF1 phosphorylation
process~\cite{rieger2005mathematical}.  These modeling choices
contribute to increase the model complexity, leading to a
difficulty to extract simple mechanisms that explain the observed
dynamics of HSR.  Complex models are also more difficult to link
with other regulation networks such as the cell cycle.
 
The scope of this paper is to develop a mathematical model of the HSR
describing the available data sets for HeLa cells for both kinetics of
HSF1 activity and cell viability under heat shock.  This model is
based on chemical kinetics laws, decreasing its dimensionality without
altering the biological interpretation of the model dynamics. The
reaction rates of HSF1 activity are estimated from Abravaya's
experimental results~\cite{abravaya1991attenuation}.  By performing a
steady-state analysis of the network, we highlight three different
 stress regimes qualified as normal, acute, and chronic,
where normal stresses correspond to pseudo thermal adaption. The
boundaries of the different regimes are conserved through the
parameter optimization process suggesting that they are highly
supported by the experimental data of~\cite{abravaya1991attenuation}.
A phenomenological link between the HSR and the cell viability reveals
that our model reproduces quantitatively the experimental data
of~\cite{gerner1976transient} on the cell viability and gives a
qualitative description of the thermotolerancy effect.
 
The modeling choices, which are deliberately simplistic, highlight the
key mechanistic ingredients of HSR. The
available data in the literature are described by a genetic network
whose guiding principle is the titration of HSF1 by chaperones. 
In this model, the
balance between renaturation and degradation of misfolded proteins is regulated 
by  the titration of MFPs by chaperones. 
This simplified description of HSRN allows the coupling with other genetic networks and may give insight in the field of cancer therapy.

\section*{Materials and Methods} 
 
\subsection*{Mathematical model of the heat shock response} 
\begin{figure}[htb] 
  \centering 
  \includegraphics[width=0.8\columnwidth]{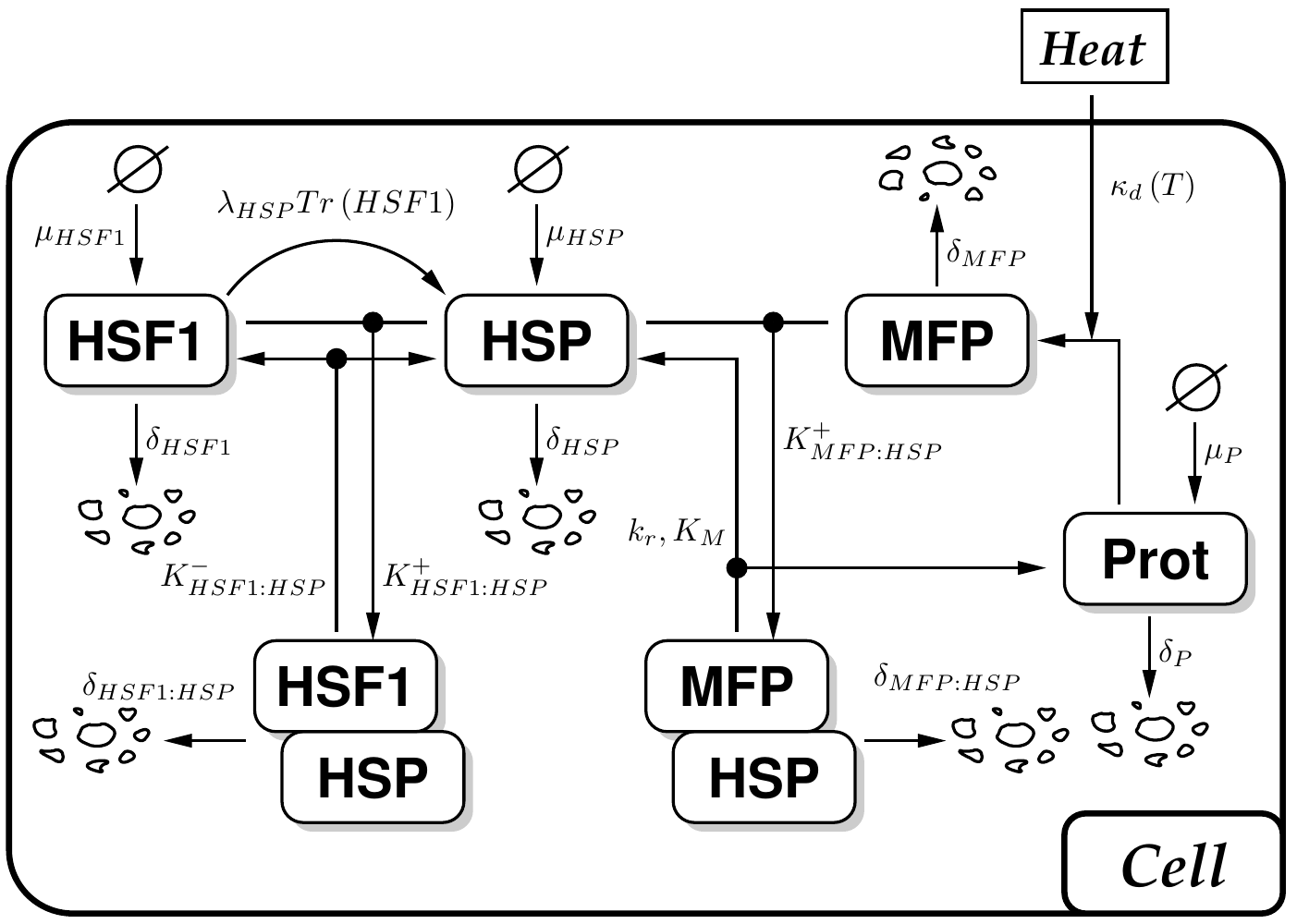}\hfill 
  \caption{\textbf{Minimal description of the Heat Shock Response Network (HSRN)}. The dynamics of this HSRN relies on the competition between 
($\mathrm{HSF1}:\mathrm{HSP}$), and ($\mathrm{MFP}:\mathrm{HSP}$). The fact that the binding affinity of 
the two complexes differs by several orders of magnitude induces the 
prevalence of refolding complex ($\mathrm{MFP}:\mathrm{HSP}$) against 
sequestration ($\mathrm{HSF1}:\mathrm{HSP}$). In this network $\mathrm{HSF1}$ is titrated by $\mathrm{HSP}$ and the protein triage governs the fate of $\mathrm{MFP}$. } 
  \label{fig:Model} 
\end{figure}

The  minimal description of the HSR genetic network (Fig.~\ref{fig:Model}) involves the transcription factor \textit{Heat Shock Factor 1} (HSF1), the chaperone proteins \textit{Heat Shock Protein} (HSP), the sequestration complex (HSF1:HSP), the  \textit{Misfolded Protein} (MFP) the chaperone complex (MFP:HSP), and the pool of protein (P). Besides the constitutive transcription rates for HSF1, HSP, and P,  
a generic function $Tr\left( HSF1  \right)$ stands for the $\left[ \mathrm{HSF1}_3:HSE \right]$ and describes the HSF1 activated transcription of \textit{HSP}. 
In this article HSP is a generic name for \textit{70 kilodalton heat shock proteins} (HSP70).  

The denaturation rate $\kappa_d\left(T\right)$, is approximated in the range 37--45\degC~range  from ~\cite{peper1998mathematical} by: 
\begin{equation} 
  \label{eq:denat} 
  \kappa_d\left(T\right)=k_d\;\left(1-0.4e^{37-T}\right)\;1.4^{T-37} 
\end{equation} 
where $T$ is the temperature in \degC. 
The denaturation rate is here the only input pathway of temperature in the network.  The protein refolding process is 
 described by a Michaelis--Menten 
 kinetics to stand for limited energetic resources.

The model of Fig.~\ref{fig:Model} is minimal as it does not take into 
account for (1) HSF1 trimerization and phosphorylation dynamics (2) 
the mRNA dynamics (3) HSF1 binding dynamics to the HSE, which are 
assumed to be fast.  These restrictions are justified because the kinetics investigated here correspond to a temperature change of several hours.  The list of involved chemical reactions is displayed in Tab~S1 of the Supplementary material.

The mathematical description by mass action law of the minimal network displayed in Fig.~\ref{fig:Model} leads to a six dimensional system of coupled differential equations. The basic reactions (transcription, degradation, reversible dimerization, the denaturation and refolding process) are implemented in the standard fashion.  
To simplify the mathematical expression the notation is compacted as follows: $x$ stands for HSF1, $y$ for HSP, $z$ for MFP, and $p$ for P. The equations  are given by:      
  \begin{subequations} 
    \label{eq:Model} 
   \begin{eqnarray} 
     \frac{d}{dt}[p] & = & \delta_p\,\left( P_T-p\right) -\kappa_{d}\left(T\right)\,\frac{p}{P_T}+k_{r}\frac{[y:z]}{K_M+[y:z]};\\ 
     \frac{d}{dt}[z] & = & \kappa_{d}\left(T\right)\,\frac{p}{P_T}-K^+_{y:z}[y]\cdot [z]-\delta_{z}[z];\\ 
     \frac{d}{dt}[y:z] & = & K^+_{y:z}[y]\cdot [z]-k_{r}\frac{[y:z]}{K_M+[y:z]}-\delta_{y:z}[y:z];\\ 
     \frac{d}{dt}[x] & = & \mu_{x}-\delta_{x}[x]-K^+_{x:y}[x]\cdot [y]+K^-_{x:y}[x:y];\\ 
     \frac{d}{dt}[y] & = & \mu_{y}+\lambda_{y} Tr\left( [x]  \right)-\delta_{y}[y]-K^+_{x:y}[x]\cdot [y]+K^-_{x:y}[x:y] \nonumber\\ & &-K^+_{y:z}[y]\cdot [z]+k_{r}\frac{[y:z]}{K_M+[y:z]};\\ 
     \frac{d}{dt}[x:y] & = & K^+_{x:y}[x]\cdot [y]-K^-_{x:y}[x:y]-\delta_{x:y}[x:y]. 
  \end{eqnarray} 
  \end{subequations} 
Despite the performed adiabatic eliminations of fast variables, 
  the mathematical model guarantees positive values for the concentrations, and thus remains biologically significant.

In Eq.~\ref{eq:Model}, the parameters $\delta_u$ are the linear degradation rates, 
$K^{\pm}_u$ the kinetic constant for heterodimerization, $\mu_u$ the 
basal transcription rates ($u$ can refer to any chemical species).   
In order to establish numerical results, a realistic sigmoidale transcription function   
\begin{equation} 
  \label{eq:TR} 
  Tr\left(\mathrm{HSF1} \right)=\frac{\mathrm{HSF1}^3}{P_0^3+\mathrm{HSF1}^3 },  
\end{equation} 
is used to account for the regulation of the Heat shock element (HSE) by an HSF1 homotrimer. $P_0$ defines the threshold of regulation and $\mu_y+\lambda_y$ is the maximal transcription rate of $HSP$. The transcription 
function Eq.~\ref{eq:TR} implicitly assumes a pseudo equilibrium for 
[HSF1], $[\mathrm{HSF1}_3]$, and $[\mathrm{HSF1}_3:\mathrm{HSE}]$. 

\bigskip 
To illustrate the displacement of the equilibrium point with the temperature, one can define for each chemical species the relative variation 
\begin{equation} 
  \label{eq:rel_ss} 
  \sigma_{[U]}\left(T\right)=\frac{[U]^*_T}{[U]^*_{37^oC}},  
\end{equation} 
which scales the value of the steady state at a given temperature $T$ to the one at 37\degC~($U$ can refer to any chemical species). 
It is worth noting that if the degradation rates of the two complexes $\mathrm{MFP}:\mathrm{HSP}$ and $\mathrm{HSF1}:\mathrm{HSP}$ vanish, then the concentrations of $\mathrm{MFP}:\mathrm{HSF1}$, $\mathrm{HSP}$, and $\mathrm{HSF1}:\mathrm{HSP}$ in steady state are independent of the temperature.

\subsection*{Mathematical model of the cell viability} 
Although the architecture of the heat shock response network is
simple, its crosstalk with the cell cycle is multiple and requires a
full modelization of the cell cycle network, which is beyond the scope
of this paper; therefore a phenomenological modelization of the
interaction between the two networks is used.  Cells are assumed to be
either in a growing state where the division occurs, or in a non
growing state where no division is possible. The transition rate from
a growing to a non growing state is assumed to be simply proportional
to the fraction of MFPs, modeling lethal effect of the lack of
functional proteins. 
As long as the proteins are misfolded (either in free form or in complex form with HSP) they are assumed to be not functional, and  therefore, they participate in the decrease of the cell survival.
In this framework, the survival probability follows an
exponential law of parameter proportional to the integral of the total
amount of MFPs within the cell over time.  The survival probability
$P$ reads:
\begin{equation} 
  \label{eq:SF} 
  P(t)=\exp\left(-\alpha\;\int_0^t \left( [\mathrm{MFP}](u)+[\mathrm{MFP}:\mathrm{HSP}](u) \right)\;du  \right), 
\end{equation} 
where $[0,t]$ is the time interval of the experiment and $\alpha$ a constant factor, independent of the thermal protocols.  
 
The survival fraction of cells, under a thermal protocol, monitors the fraction of cell in a colony for which the growing rate is not affected by the heat shock treatment (see~\cite{gerner1976transient} for detailed protocols). In the framework of the probability (Eq.~\ref{eq:SF}), the survival fraction corresponds to the ratio  between the survival probability for the given protocol to the survival probability  at a constant 37\degC~temperature for the same duration.

\subsection*{Experimental data} 
The experimental data used to estimate the parameters of the network 
are taken from the Fig. 8-A of~\cite{abravaya1991attenuation}. The 
heat shock experiments are performed on HeLa S3 cells with a water bath. 
Abravaya \textit{et al.}~\cite{abravaya1991attenuation} measured by run-on assay the kinetics of activated HSF1 in Hela cells grown at a temperature of 37\degC~ and submitted to temperature increase up to 41\degC, 42\degC, and 43\degC~during 4 hours. Activated HSF1 corresponds to phosphorylated trimer of HSF1. In the framework of the model (Fig.~\ref{fig:Model}), activated HSF1  is proportional to $[\mathrm{HSF1}]^3$ because phosphorylation and trimerization are assumed to be rapid processes. 
 
The experimental data of survival response of HeLa cells are taken 
from the figures 1--3 of~\cite{gerner1976transient}. 
The thermal protocol consists on  measuring the survival fraction of cells after a  heat shock between 30~min and 4~hours and a temperature from 41$^\circ$C to 45$^\circ$C.

\subsection*{Adjustment and goodness of fit} 
To measure the goodness of fit for a given parameter set, we have defined a root mean square (RMS) error between the experimental measures of activated HSF1 and  the numerically computed $[\mathrm{HSF1}]^3$. At each fitness computation, the scale factors have been adjusted to minimize the RMS error over the three heat shock experiments. The pool of proteins is set to 4.5~mM by fixing $P_T=4500$ (\cite{Milo2013}). The half--life of the native proteins is set to 10~h ($\delta_T=0.069$). Setting these parameter scales the concentration in $\mu$M unit and the time in hours. 

Adjustment has been carried out by using a nonlinear optimization procedure based on a modified Levenberg--Marquardt algorithm (MINPACK)  
software suite~\cite{more:_minpac}.  
The non linear optimization procedure has been initiated by random value for the parameters. Since the order of magnitude of the protein half--life is well known, the degradation rates  have been randomly chosen to ensure a half--life in the range of 6--30 hours and have not been included in the non linear optimization.      
Numerical integration of 
ordinary differential equations has been performed with the SEULEX 
algorithm~\cite{hairer96:_ODE} which is well adapted to stiff systems. 
The convergence of the adjustment has been monitored by verifying that the optimum has been repeatedly reached.


\section*{Results}

\subsection*{Parameters estimation from experimental data} 

The model investigated in this paper is made of the regulation network displayed in Fig.~\ref{fig:Model} as well as the phenomenological definition of the survival probability Eq.~\ref{eq:SF}. In the following it will be referred as the Heat Shock Response Network (HSRN).

\begin{figure}[htb] 
  \centering 
    \includegraphics[width=0.5\columnwidth]{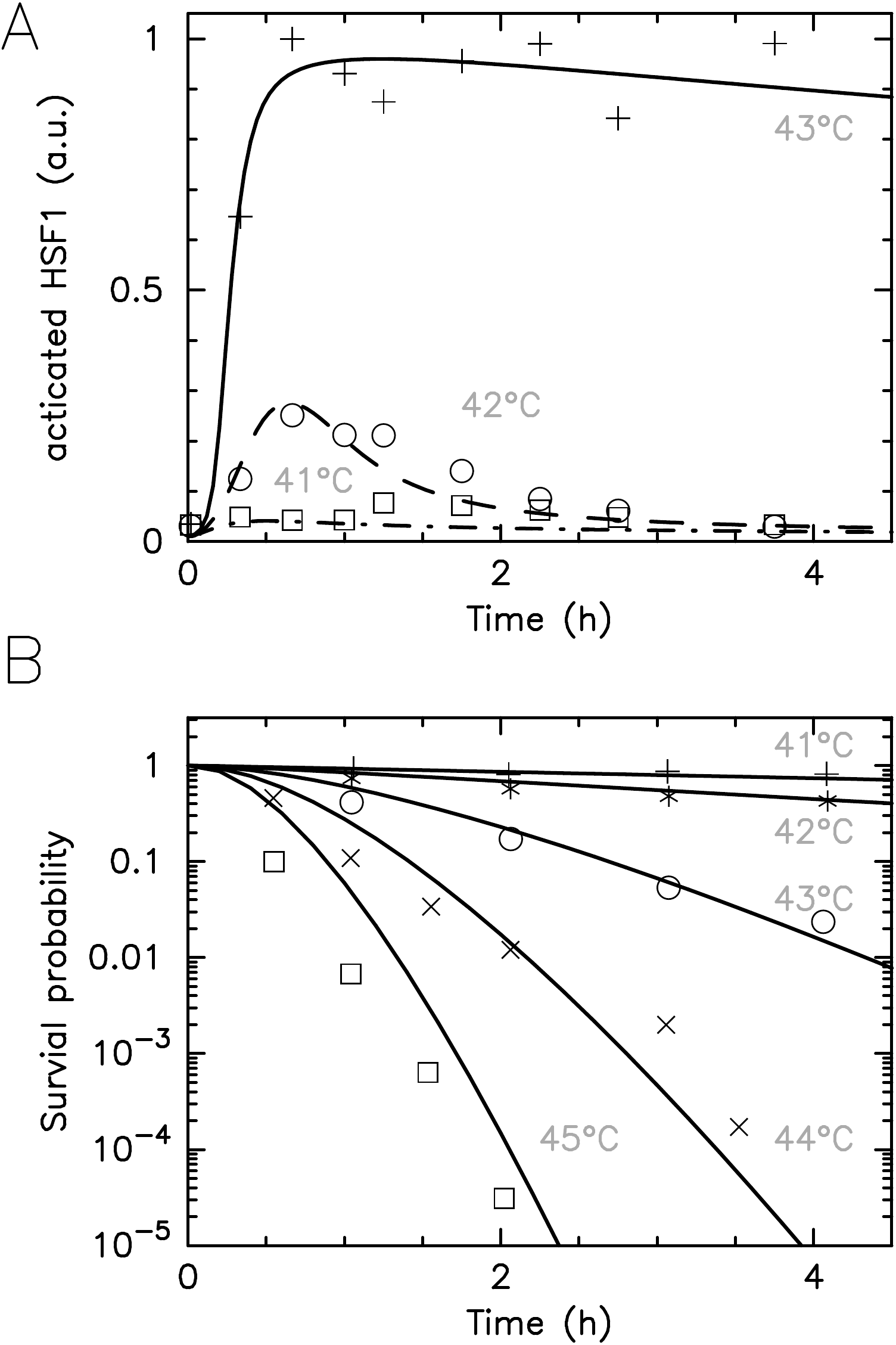} 
 
  \caption{{\bf Adjustment of HSRN on experimental data on Hela cells.}   
(A) Parameters estimation based on continuous heat shock data set from~\cite{abravaya1991heat} displayed by the squares, circles, and crosses for a heat shock of 41\degC, 42\degC, and 43\degC~respectively. Lines are the output of the model for the variable $[\textrm{HSF1}]^3$ which stands for activated HSF1: mixed, dashed, and solid lines correspond to a temperature of 41\degC, 42\degC, and 43\degC~respectively. At initial time, the network is in the steady state at 37\degC. 
(B) Survival fraction of cells exposed continuously to increased  temperatures for different durations. Data are  taken from~\cite{gerner1976transient} and re-plotted as points, errors bars have been omitted for clarity, the heat shock amplitude is indicated directly on the figure. The continuous lines are the results of the HSRN.  
} 
  \label{fig:HS_PE} 
\end{figure}

The HSRN implies kinetic parameters whose values are unknown.  The
parameter values are adjusted to reproduce, at best, two complementary
data sets for Hela cells (See Methods for details). The first set of
experiments is chosen to monitor is the HSF-induced transcription of
protein chaperones under continuous heat
shock~\cite{abravaya1991attenuation}.  The second quantifies the
survival fraction as a function of the duration and intensity of the
heat shock~\cite{gerner1976transient}.  Both experimental results are
re-plotted in Fig.~\ref{fig:HS_PE}.

The two experimental data sets displayed in Figure~\ref{fig:HS_PE}), independently reveal  a sharp transition around 42\degC.
For a temperature under 41\degC, the heat shock response is almost undetectable with the available time lapses; whereas above 43\degC~ the chaperone transcription remains at a high value while the survival probability decreases exponentially with the exposure time (3 hours at 43\degC~leads to a division by 10 of the growing rate).   
Reproduction of this sharp transition is  a challenge for modeling because the temperature input is a smooth function.

The HSRN  describes  quantitatively the experimental data sets for both HSF1 kinetics of activation and cell viability under heat shock (Fig.~\ref{fig:HS_PE}). 
Regarding the chaperone transcription (Fig.~\ref{fig:HS_PE}-A), the overshoot at 41\degC~and 42\degC~is well 
captured as well as the saturation at 43\degC, while the respective 
levels are in very good quantitative agreement.  
Meanwhile, the obtained survival probability is consistent with experimental data (Fig.~\ref{fig:HS_PE}-B) with main discrepancies arising from short heat shock of high intensity.    

\begin{table}[htb] 
  \centering 
  \caption{Estimated parameters for the HSRN} 
  \label{tab:par} 
  
 \begin{tabular}[htb]{p{3cm}p{2cm}p{7cm}p{2cm}} 
 Parameter&unit&description&value\\ \hline\hline 
$\ln(2)/\delta_{HSF1}$&(h)&HSF1 half--life&        6.62\\
$\ln(2)/\delta_{HSP}$&(h)&HSP half--life&       13.33\\
$\ln(2)/\delta_{HSF1:HSP}$&(h)&HSF1:HSP half--life&        7.94\\
$\ln(2)/\delta_{MFP:HSP}$&(h)&MFP:HSP half--life&       30.00\\
$\ln(2)/\delta_{MFP}$&(h)&MFP half--life&        3.18\\
$\mu_{HSF1}$&($\mu$M.h$.^{-1}$)&HSF1 basal transcription rate&        3.41E-03\\
$\mu_{HSP}$&($\mu$M.h$.^{-1}$)&HSP basal transcription rate&        7.51E-12\\
$\lambda_{HSP}$&($\mu$M.h$.^{-1}$)&HSP active transcription rate&        1.33\\
$P_0$&($\mu$M)&HSP transcription regulation threshold&       24.85E-03\\
$K^+_{HSF1:HSP}$&($\mu$M$.^{-1}$.h$.^{-1}$)&HSP:HSF1 binding affinity&       61.40\\
$K^-_{HSF1:HSP}$&(h$.^{-1}$)&HSP:HSF1 unbinding rate&       15.29\\
$K^+_{MFP:HSP}$&($\mu$M$.^{-1}$.h$.^{-1}$)&MFP:HSP binding affinity&      827.58\\
$k_d$&($\mu$M.h$.^{-1}$)&denaturation  rate&        1.68\\
$k_r$&($\mu$M.h$.^{-1}$)&maximal renaturation  rate&       11.89\\
$K_M$&($\mu$M)&renaturation Michaelis constant &      0.411\\
$\alpha$&($\mu$M$.^{-1}$.h$.^{-1}$)&death rate &      0.171\\
 \end{tabular} 
\end{table}

As usual, the parameters estimation does not provide a unique
parameter set.  To quantify the dispersion of the parameter sets, the parameters values are plotted, for the 100 best parameters sets
obtained, as a function of their fitness normalized to the best one 
(Fig.~S1).  The wide dispersion of the parameters over the different
sets is not a surprise due to the lack of experimental data
precision. However, it appears that all the obtained sets are in the
same region of the parameter space, meaning that the best set
reproduced in Tab.~\ref{tab:par} is representative of all the
solutions. 

The steady state at  37\degC~is consistent with the
biological attempts.  Indeed, at 37\degC~almost all the HSF1 are in a complex
form with HSP whereas HSP proteins are mainly in the monomer form and
thus available for the folding of newly synthesized proteins
(Tab.~\ref{tab:par_ss}). The concentration of MFPs is low and most of
them are in a complex with $\textrm{HSP}$. Once again, these results
are fairly well conserved over the optimization process (Fig.~S2)
without any requirement on the steady state.  In particular, the total
concentration for HSF1 and HSP at 37\degC~are in good quantitative
agreement with the experimental measured value of 0.03~$\mu$M for HSF1
and 1~$\mu$M for HSP70~\cite{nagaraj2011deep}.

\begin{table}[htb] 
  \centering 
  \caption{Estimated steady state at 37\degC} 
  \label{tab:par_ss} 
 \begin{tabular}[htb]{p{3cm}p{3cm}p{1cm}} 
  Species&value&unit\\ \hline\hline  
$[HSF1]^*$&        8.36$\times 10^{-3}$&($\mu$M)\\
$[HSP]^*$&      0.871&($\mu$M)\\
$[HSF1:HSP]^*$&       0.029&($\mu$M)\\
$[MFP:HSP]^*$&       0.038&($\mu$M)\\
$[MFP]^*$&        1.40$\times 10^{-3}$&($\mu$M)\\
$\Sigma[HSF1]^*$&       0.037&($\mu$M)\\
$\Sigma[HSP]^*$&      0.938&($\mu$M)\\
$\Sigma[MFP]^*$&       0.039&($\mu$M)
 \end{tabular} 
 \end{table}

\subsection*{Detailed kinetics for continuous heat shock}

\begin{figure}[htb] 
  \centering 
  \includegraphics[width=\columnwidth]{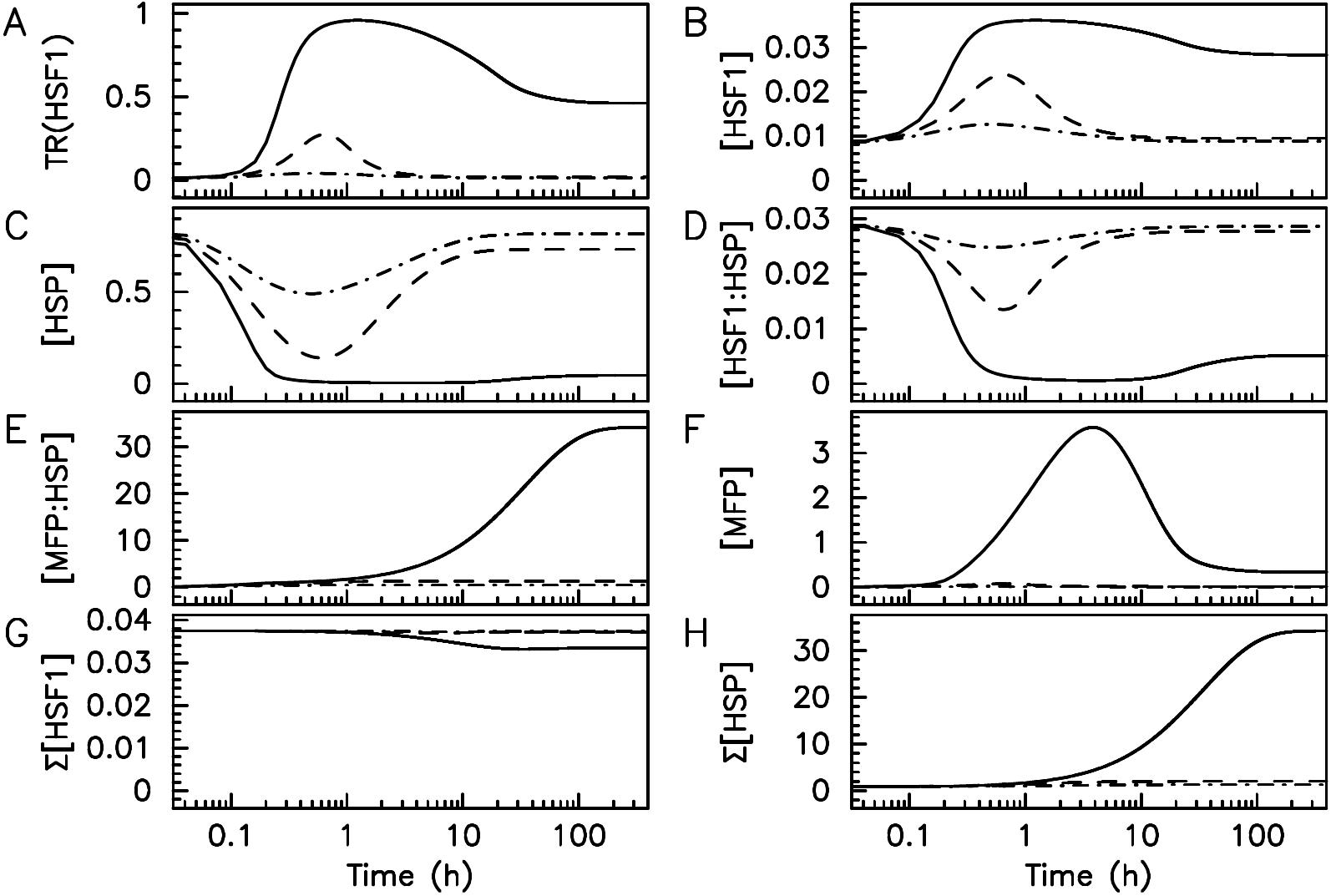} 
  \caption{{\bf Detailed chemical kinetics of the heat shock response to continuous heat shock.} Time evolution of HSP transcription rate (A) and of various concentrations (B-H) for a continuous heat shock at 43\degC~(solid), 42\degC~(dashed), and  41\degC~(dashed-doted). Before time zero, the system is in equilibrium at a 37\degC~temperature. } 
  \label{fig:HS_all} 
\end{figure}

To give a clear understanding of the HSRN dynamics, Fig.~\ref{fig:HS_all} presents the time evolution of concentrations until the new equilibrium state is reached. 
Let us first describe the strongest heat shock at 43\degC.  During the
first ten minutes, all free HSP monomers bind with created MFPs
(Fig.~\ref{fig:HS_all}-C). Then the $\mathrm{HSF1}:\mathrm{HSP}$
complex starts to dissociate to free HSP monomers
(Fig.~\ref{fig:HS_all}-D), increasing simultaneously $\mathrm{HSF1}$
monomers (Fig.~\ref{fig:HS_all}-B) which rapidly self trimerize and
activate the $HSP$ transcription (Fig.~\ref{fig:HS_all}-A). The
transcription rate reaches its maximum after one hour.  The adaptation
of the HSP concentration to the heat condition is not instantaneous
due to the long life time of HSP (Fig.~\ref{fig:HS_all}-H).  As a
consequence, a sudden increase of temperature from 37\degC~to
43\degC~induces a transient accumulation of monomeric MFP
(Fig.~\ref{fig:HS_all}-F).  To catch up, the $HSP$ transcription
overshoots its stationary value until the MFP are all chaperoned and
then relaxed to their stationary values.

Once the transcription rate of HSP is sufficient to bind all the heat 
induced MFPs (around 10 hours after the heat shock beginning), free 
monomeric HSP are available for a binding with HSF1 monomeric forms 
repressing the expression of its own gene. The total concentration of HSP therefore increases with smaller rate.  The free MFPs are kept to a low value 
and all the concentration uniformly relax to the new steady state.

For the weaker heat shock, the heat induced MFPs are 
binded  by the free monomeric HSP without requiring an unbinding of the entire 
 pool of $\mathrm{HSF1}:\mathrm{HSP}$ heterodimers  (Fig.~\ref{fig:HS_all}-D). The transcription of $HSP$ 
increases slightly due to the increase of monomeric HSF1 but not in 
the same proportion as for a 43\degC~heat shock. After the overshoot, the 
monomeric HF1 concentration relaxes.

\subsection*{Identification of three HSRN working regimes} 
\begin{figure}[!ht] 
  \centering 
  \includegraphics[width=0.75\columnwidth]{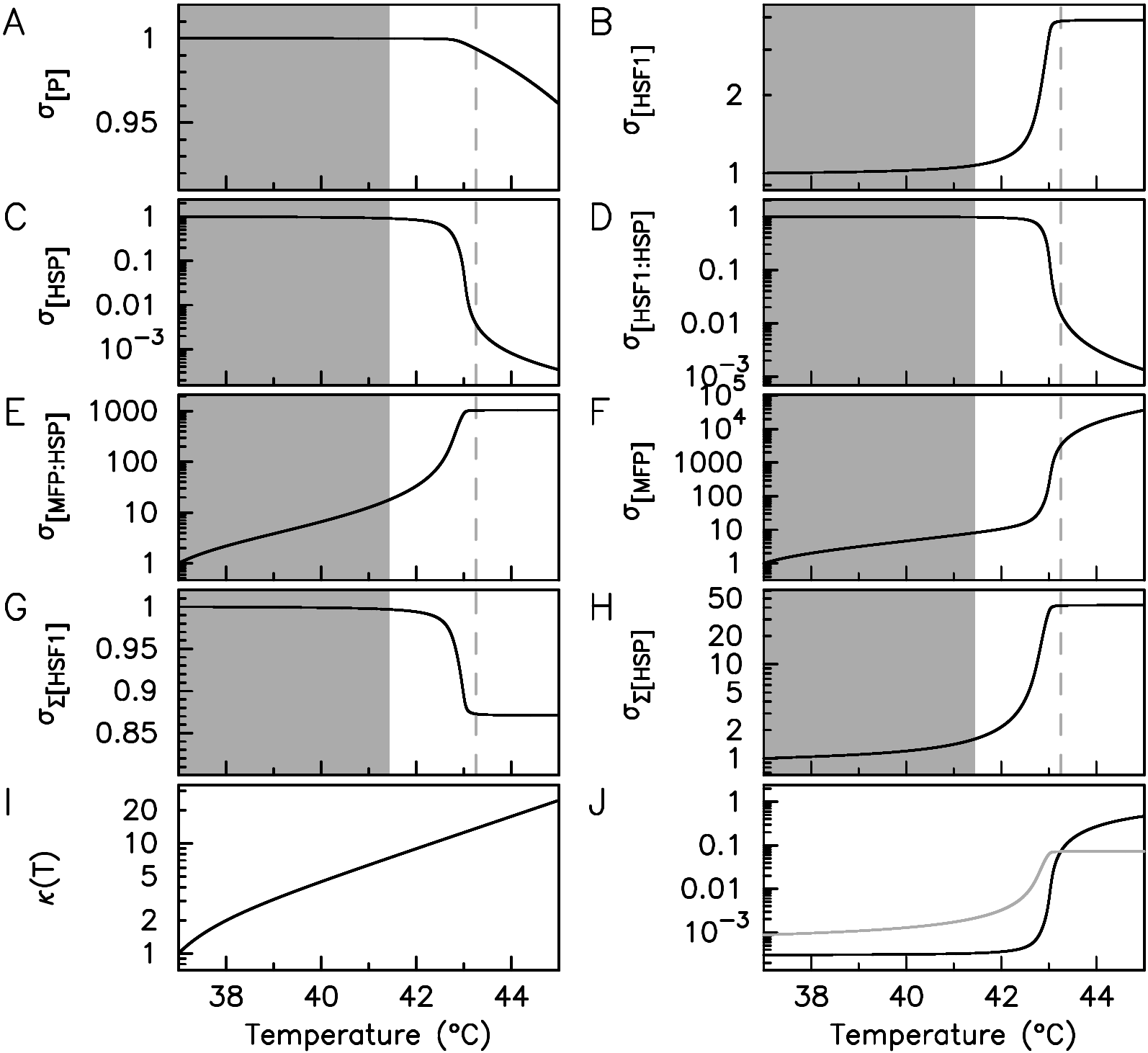} 
  \caption{\textbf{Pseudo thermal adaptation and temperature 
      break-up.}  A-H Numerical computation of the relative variations $\sigma$ of the HSRN steady states induced by a temperature increasing. $\Sigma$[HSF1] (resp. $\Sigma$[HSFP]) stands for the total 
    concentration of HSF1 (resp. HSP). 
The relative variation is defined as the 
    ratio between the steady state at a temperature $T$ to the one 
    at $37^oC$ (see MM for details). 
The shaded area highlight the pseudo thermal adaptation temperature range, whereas the grey dash line indicates the break-up temperature ($T=T_{bk}$).  
 I Temperature dependency of used  denaturation function (Eq.~\ref{eq:denat}).  
J Flow balance  of denaturated proteins management $\eta_1$ (black) and $\eta_2$ (grey).} 
  \label{fig:HS_PTF} 
\end{figure}

All the experimental data highlight a sharp transition in the response 
around 42$^\circ$C. Beyond this threshold, activated HSF1 raises to a 
constant maximum value and the cell viability quickly decreases.  As 
mentioned previously the HSRN provides a good quantitative description of this sharp 
transition even if a smooth function is used as temperature input (Fig.~\ref{fig:HS_PTF}-I).

The different working regimes of the HSRN are characterized by a
steady state analysis. The concentration in equilibrium at a given
temperature is numerically computable because the HSRN includes
protein synthesis and degradation.  It is worth noting that the HSRN
has always only one stable fixed point; therefore, the information
extracted from the steady state analysis concerns the concentration
in equilibrium at a given temperature only.  However, the variation of
the equilibrium concentration with respect to the temperature is
sufficient to give insight on the dynamical behavior of the HSRN.
 
In many models of the literature, the degradation of dimers are not
included, leading to drastically changes in the behavior of the steady
state with the temperature. In this approximation, the HSF1 steady
state concentration becomes invariant with the temperature, which can be 
interpreted as a mathematical signature of a thermal adaptation of the
HSR.  As the steady state analysis is independent of the adiabatic
elimination of fast variables, we affirm that the thermal adaptation
is impossible to achieve with this chemical reaction network, without
contradiction such as an infinite lifetime of a protein complex.

The Figure~\ref{fig:HS_PTF} displays the steady state of the HSRN for
a wide range of temperature for parameters set of Tab.~\ref{tab:par}.
A first result is that almost all proteins keep their correct
conformations in steady state up to a temperature of
45\degC~(Fig.~\ref{fig:HS_PTF}-A). The variation of the concentration
of native proteins is less than 5\%. In a simpler picture, the concentration of 
native protein  can be considered as a constant value,
whereas the misfolded proteins are created with a constant flux
$\kappa_d(T)$.

On the opposite, the concentrations of the core players of the HSRN
display a sharp variation with temperature at equilibrium
(Fig.~\ref{fig:HS_PTF}-B-H). To quantify these sharp variations, one
can define two temperature thresholds, named thereafter $T_{Th}$ and
$T_{Bk}$. $T_{Th}$ characterizes the beginning of the sharp increase,
and $T_{Bk}$ the achievement of the maximum of HSP transcription.
Surprisingly, the values of the two temperature thresholds
$T_{Th}\simeq41.5^\circ C$ and $T_{Bk}\simeq 43.3^\circ C$ are highly
conserved over the best optimized parameters set (see Fig.~S2 of the
Supporting Material).
 
A temperature increase below $T_{Th}\simeq42^\circ C$ induces a weak 
variation of the steady state concentration. As a criterion, one can 
define $T_{Th}$ as the limit temperature above which the steady state 
has a relative variation less than 10\% for the central HSR network 
partners HFS1, HSP, and $\mathrm{HSF1}:\mathrm{HSP}$ (see Materials and Methods). 
For a continuous temperature increase below $T_{Th}$, the heat 
induced misfolded proteins are buffered by the chaperone monomers 
without inducing a significant increase of HSP transcription.  In this 
operating temperature, the network exhibits a pseudo thermal 
adaptation, even with a non vanishing degradation rate of the proteins 
complex. 
 
A break up in the dynamical behavior of the HSRN arises for $T>T_{Bk}$. 
When the temperature increase exceeds $T_{Th}$, the initially 
available concentration of chaperone monomers is not sufficient to refold 
the denaturated proteins, then the complex 
$\mathrm{HSF1}:\mathrm{HSP}$ dissociates and the transcription of $HSP$ 
is thus activated.  In this operating regime, the transcription rate of HSP 
takes its maximum value but is still not fast enough to dominate the heat created MFPs by 
binding. To quantify this transition, on can define $T_{BK}$ as the temperature for which the HSF1 induced transcription of HSP raises 99\% of its maximum  value $\lambda_{\textrm{HSP}}$.  
In this regime,  misfolded proteins 
are then rather degraded than refolded by HSP, as it will be explain in details further. 
 
A mathematical estimation for the break-up temperature $T_{Bk}$ arises 
by equating the denaturation flux $\kappa_d(T)$ to the sum of the 
maximal flux of newly synthesized 
$\mu_{\textrm{HSP}}+\lambda_{\textrm{HSP}}$ and the maximal flux of 
renaturation $k_r$ \textit{i.e.} 
$\kappa_d(T_{BK})=\mu_{\textrm{HSP}}+\lambda_{\textrm{HSP}} + k_r$. 
Using the approximation $\kappa_d(T_{BK})=k_d\;1.4^{T_{BK}-37}$ and 
the parameters of Tab.~\ref{tab:par}, a $T_{BK}=43.12^\circ$C is found out and is in good agreement 
with numerical result. 
 
\subsection*{Triage of MFP} 

The key role of the heat shock response is to manage the misfolded
proteins, through degradation, sequestration or refolding
processes~\cite{garrido2010heat}.  For instance, the HSRN involves
different pathways to manage denatured proteins either through direct
degradation of denatured proteins with a constant rate
$\delta_{\textrm{MFP}}$, or through complexation with chaperones. The
chaperone complex $\textrm{MFP}:\textrm{HSP}$ can either be degraded
or renatured.  The degradation occurs at a constant rate
$\delta_{\textrm{MFP:HSP}}$, implying the destruction of the denatured
proteins and chaperones. The rate of the renaturation process is given by
$k_r/(K_M+\textrm{[HSP]})$. In this latter case, the chaperone is
released from the complex, and is available for complexation with
other misfolded proteins.
 
The balance between the different pathways in
the MFP triage is characterized by two flux balance indexes:
\begin{equation} 
  \label{eq:flux_bal} 
  \eta_1=\frac{\delta_{\textrm{MFP}}}{\delta_{\textrm{MFP}}+K^+_{\textrm{[MFP:HSP]}}\textrm{[HSP]}}; \quad \eta_2=\frac{\delta_{\textrm{MFP:HSP}}}{\delta_{\textrm{MFP:HSP}}+k_r/(K_M+\textrm{[MFP:HSP]})};  
\end{equation} 
both taken value in $[0,1]$. A low value of $\eta_1$ (resp. $\eta_2$) 
indicates a prevalence of $\textrm{MFP:HSP}$ complexation on 
$\textrm{MFP}$ degradation, whereas a low value of $\eta_2$ indicates a prevalence of renaturation on 
$\textrm{MFP:HSP}$ degradation.  

For weak thermal stresses ($T<T_{BK}$), both $\eta_1$ and $\eta_2$
remains less than 10\% and the renaturation process is dominant
(Fig.~\ref{fig:HS_PTF}-J).  Whereas beyond $T_{BK}$, $\eta_1$ raises a
value close to unity while $\eta_2$ remains to a constant value of
0.12. This implicates a prevalence of the $\textrm{MFP}$ degradation
on the complexation, due to the lack of free available $\textrm{HSP}$.
If renaturation process is the dominant pathway for moderate stresses,
for acute stresses the degradation of $\textrm{MFP}$ dominates, and 
leads to a loss of functional protein in cell
(Fig.~\ref{fig:HS_PTF}-A).

\subsection*{Cell viability increased by dose fractionation} 
\begin{figure}[htb] 
  \centering 
  \includegraphics[width=0.5\linewidth]{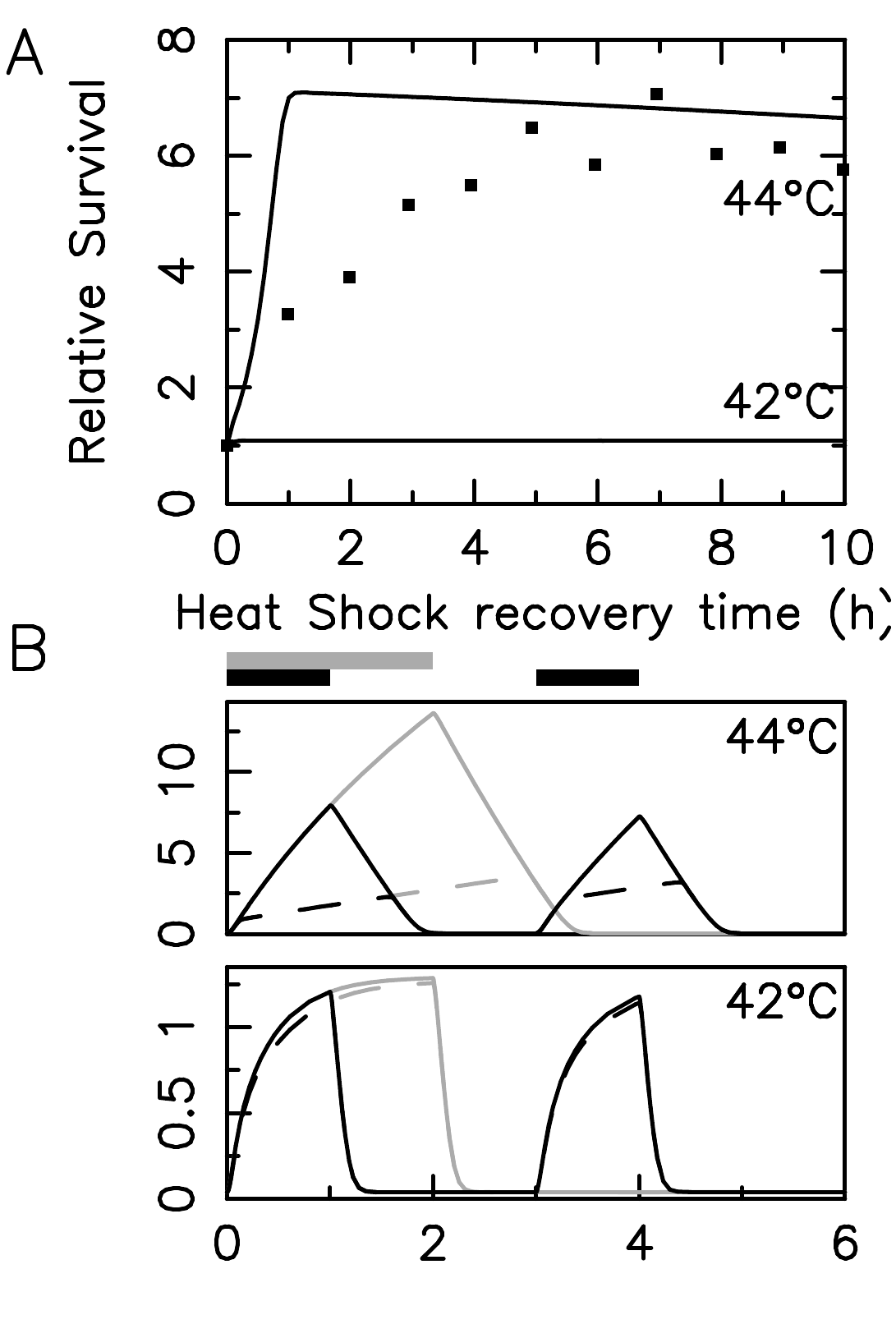} 
  \caption{{\bf Cell survival upon  fractionated  heat shock.}  
(A) Relative survival fraction for two one-hour heat shock separated  by a given time recovery time at 37\degC~for the indicated temperature increase ( 42 and 44\degC).
(B) Detailed kinetics of the MFP occupancy, for a two hours heat shock (grey) and two one-hour heat shock separated by a two-hours recovery at 37\degC~(black). Continuous lines stand for the total MFP concentration kinetics ($\left[\textrm{MFP}+\textrm{MFP:HSP} \right]$), dashed lines for the chaperoned misfolded protein kinetics ($\left[\textrm{MFP:HSP}\right]$).   
} 
  \label{fig:HS_BI} 
\end{figure}

Setting a thermal lethal dose is complex as it can not be inferred 
from the measurements of cellular viability under continuous thermal 
shock. Indeed, it is well known that the fractionation of the exposure time 
to the temperature rise induces an increase in cell 
survival~\cite{gerner1976transient}.  
For example, a thermal protocol made of two one-hour heat shock at 44$^\circ$C~ separated by two hours recovery at 37$^\circ$C,  increases the viability by six fold as compared to a two-hours treatment (Fig.~\ref{fig:HS_BI}-A).  The relative survival is then 
defined as the ratio between the survival fraction for a given 
recovery time at 37$^\circ$C and those for a zero recovery time.

The HSRN describes the survival increase due to fractionation (Fig.~\ref{fig:HS_BI}-A).  
Applying similar protocols (two one hours heat shock with varying recovery time) on the HSRN results in a good quantitative agreement with experimental data for a survival probability.  
The main discrepancy arises from the 
recovery time less than 2 hours but the asymptotic value is correct. 
The minimal model is then fully able to explain the experimentally 
observed viability increase.  
If we apply the same fractionation protocols 
for various heat shock amplitudes, one find that the 
survival probability remains constant for heat stress lower than 
42$^\circ$C. Beyond 42$^\circ$C the  relative survival  probability increases rapidly with the heat increase (data not shown). 
 
To highlight the effect of dose fractionation, (Fig.~\ref{fig:HS_BI}-B) displays the concentration if MFPs with and without recovery  for a two hours recovery time. For a  42$^\circ$C heat stress,  all MFPs are chaperoned by HSPs 
and under recovery, the total MFPs concentration relaxes within ten 
minutes. The total MFPs concentration quickly raises (within one hour) 
its saturation value under continuous heat shock because the initial 
HSP pool is sufficient to manage MFPs and no HSP 
transcription is thus needed. 
A second heat shock induces the same dynamics and so cell viability does not vary with the time lapse between the two shocks.

On the opposite, in the case of a 44$^\circ$C heat stress, the 
initial HSPs pool is insufficient to manage the MFPs,  HSPs transcription 
is activated. Thus it takes much longer time to reach the saturation, 
then MFP accumulates in free form during heat stress. The total MFP 
concentration decreases slowly (within one hour) under recovery. 
The [MFP:HSP] kinetics under fractionation reveals that 
the second stress benefits from the HSP transcript to manage the first, 
as the fraction of MFP in complex with HSP is greater in the second 
shock than in the first.  The main effect of the fractionation 
arises from the free MFP. The recovery time is used by HSP to eliminate 
the backlog. The fractionation creates much less MFP between 3 
and 4 than the two hours heat shock between 1 and 2.

\subsection*{Predictions based on the HSRN}

The HSRN allows to study the kinetics of activated HSF1 upon continuous heat shock upon parameters modifications. One can investigate the case of an overexpression of both \textit{hsf} and \textit{hsp}, as well as the suppression of cellular function like transcription or proteasome activities by drug inhibitors.  

\begin{figure}[ht] 
  \centering 
  \includegraphics{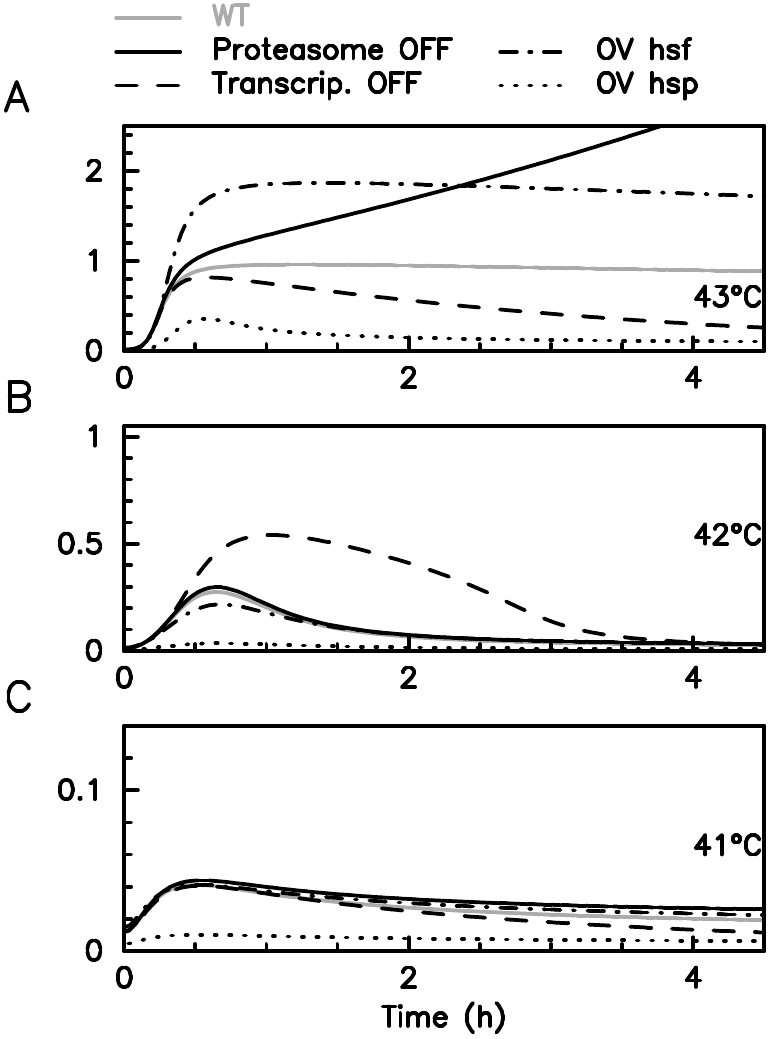} 
  \caption{\textbf{Predicted kinetics of activates HSF1 upon continuous heat shock in case of biochemical modifications.}  A-C Numerical computation of [HSF1:HSE] for a continuous heat shock of 43\degC, 42\degC, and 41\degC. Grey lines are the same results as in Fig~2 (reference experiments); black lines (resp. dashed lines) simulate a proteasome (resp. transcription) inhibition apply at $t=0$; dots line (resp. dashed dot) simulate an constitutive \textit{hsp} (resp. \textit{hsf}) overexpression.  } 
  \label{fig:HS_PRED} 
\end{figure}


As a first step, the transcription or the
proteasome activity are inhibited by a drug treatment applied
simultaneously with the continuous heat shock. To simulate a
transcription (resp. proteasome) inhibition we set at time $t=0$ the
transcription parameters $\mu$ and $\lambda$ (resp. the degradation
parameters $\delta$) to zero, and then apply an heat stress of various
intensity (43\degC, 42\degC, and 41\degC). 

As shown in Fig.~6 (solid line), blocking the proteasome does not
affect significantly the dynamical response at 42\degC, and 41\degC,
in the regime of pseudo thermal adaptation. At 43\degC, it induces a
constant increase in the activated HSF1 instead of a saturation value
found in the wild type. This is simply due to the suppression of HSF1
degradation (and then constant increase of HSF1 concentration) whereas
HSP is monopolized by misfolded proteins and then not available to form a 
complex with HSF1.

Blocking the transcription induces a faster relaxation of activated
HSF1 at 43\degC~and 41\degC~(Fig.~6-A and C dashed line) due to the
degradation of HSF1. Besides that blocking, the transcription enhances the
concentration of activated HSF1 at 42\degC~(Fig.~6-B dashed
line) due to the lack of newly synthesized HSP proteins that
should complex with HSF1 in the wild type. Newly synthesized HSP are not
induced at 41\degC~and are monopolized by MFP at 43\degC, enhancing  the activated HSF1 only at 42\degC~only.

In a second set of protocols we seek for the response in case of an
overexpression of the core player HSP and HSF. To simulate the
overexpression, we increase the transcription rate of HSF1 or HSP. To
be specific the overexpression of HSF (HSP resp.) is made by
increasing $\mu_{HSF}$ of 25\% ($\lambda_{HSP}$ of 400\% resp.). The
overexpression factor are chosen to highlight the effect in a clear
setting.  Starting from the steady state at 37\degC~with the
overexpression, we apply a heat stress of various temperatures.

In the case of an \textit{hsp} overexpression (dot lines of Fig.~6),
no response is found for an heat shock at 42\degC~or 41\degC. For a
43\degC~heat shock, the amplitude of the activated HSF1 decreases by a
4 fold.  The kinetics develops a relaxation and are quite similar to
the wild type kinetics at 42\degC. The overexpression of \textit{hsp}
also lowers the temperature threshold $T_{Th}$ and
$T_{Bk}$. Increasing the HSP levels allows to increase the buffering
capabilities of misfolded proteins in the cell

Finally the overexpression \textit{hsf} does not change significantly
the dynamical response. The main signature is founded for a 43\degC~
heat shock where the amplitude of the response is increased by a
factor 2.  At 42\degC, a slight decrease in the amplitude of the
response is found, due to the increased of HSP concentration in steady
state at 37\degC~(induced by the overexpression of \textit{hsf}) that
enhances the buffering capabilities.


\section*{Discussion}

A minimal mathematical model~ highlights the key ingredients of the 
kinetics of the heat shock response while giving an acute description 
of the experimental data. 
The core of the HSR is a competition between two complexes involving the chaperones HSP, the first one with the transcription factor  HSF1 ($\mathrm{HSF1}:\mathrm{HSP}$), and the other one with the misfolded proteins ($\mathrm{MFP}:\mathrm{HSP}$). The fact that the binding affinity of 
the two complexes differs by several orders of magnitude induces the 
prevalence of refolding complex ($\mathrm{MFP}:\mathrm{HSP}$) against 
sequestration ($\mathrm{HSF1}:\mathrm{HSP}$). 

In the modelization of the heat shock response network, the dynamics of these complexes (association and dissociation) play a key role. It is commonly found in the literature that the protein complex dissociation is mediated by a third agent \textit{e.g.} 
\begin{verbatim}
HSF:HSP + MFP -> HSF + MFP:HSP
HSF_3:HSE + HSP -> HSE + 2HSF +HSF:HSP 
\end{verbatim}
The first reaction indicates that the sequestration complex breaks up
in the presence of MFP, while the second implies a repression of HSP
directly on its promoter.  Spontaneous or activated complex
break up appears as two independent reaction pathways. 
However, the spontaneous break up must always be taken into account, due to
energetic transfer from the solvent.
  The results
developed in this paper reveal that spontaneous dissociation of the
complex is a sufficient mechanism to explain the kinetics of the heat
shock response. In particular, the fact that at 37\degC~ more than 99\%
of HSF is in complex with HSP (with vanishing MFP) may be misinterpreted
as a dissociation mediated by MFP, whereas it is not required.  Until
clear experimental evidence of any prevalence between the two molecular
pathways of dissociation, there is no reason to include a
duplicate in the modeling.

The dynamics of HSF1 trimerization and phosphorylation  are always set in details in the models of the literature. One of the messages behind the results of our work is that a fine description of the continuous heat shock kinetics does not require the inclusion of these fast dynamics. In other words, the dynamics of the HSF1 trimerization and phosphorylation are not key ingredients here.

Several previous published models have failed to describe the transition in 
the kinetics between a 42\degC~and a 43\degC~temperature increase 
(in particular the plateau in the response at a 43\degC~) even with 
detailed 
modeling~\cite{rieger2005mathematical,szymanska2009mathematical,petre2011simple}. 
It is instructive to compare the minimal model developed in this work 
with the detailed model developed by Sriram~\textit{et 
  al.}~\cite{sriram2012detailed} because both studies use the same 
experimental data for parameters estimation and have a similar 
quantitative description of the data. The model~\cite{sriram2012detailed} gives a detailed description of the HF1 trimerization, the 
hyper-phosphorylation of $\mathrm{HSF1}_3:\mathrm{HSE}$, and the 
conformal changes of newly synthesized HSP. On the other hand,  the interaction 
between HSP and misfolded proteins is not described. Moreover, the link 
between the experimental temperature increase and the level of stress 
in the model is not straightforward. Therefore, even if a complete and instructive  
bifurcation analysis is performed, the detailed model of 
~\cite{sriram2012detailed} does not provide a simple interpretation of the 
kinetics of the HSR network. In comparison, the minimal model developed here is certainly simplistic but provides (1) a similar agreement with experimental data; (2) a direct link with experimental temperature that facilitates the model prediction, and (3) a detailed kinetics of the MFPs. 

An original aspect of this work is the connection between the HSR network and
the surviving probability of the cell after heat shock.  Although the
link that is used is purely phenomenological (insofar as no direct
coupling mechanisms between the HSR and the cell cycle has been used),
this basic description reproduces effectively the variation of cell viability
to the intensity and duration of heat shock, and includes the thermotolerance effect.  
These pioneering results
in coupling between HSR and cell proliferation are thus
encouraging.

The thermotolerance in the heat shock response has been investigated in a previous  study~\cite{rybinski2013modelling} by using a previous parameters estimation~\cite{szymanska2009mathematical}. Rybinski \textit{et al.} analyze the accumulation of misfolded protein upon fractionated heat stress at 42\degC. Their results reveal a strong influence of the fractionation on the concentration of free MFP, which is in contradiction with the results presented in this paper where no influence is found at 42\degC. The key point is that in Rybinski \textit{et al.} a 42\degC~heat stress induces an increase of the HSP concentration in a rapid time scale due to the use of  a 15 min HSP half-life. Since we have restricted the HSP protein half-life in a biologically relevant range for mammals  (6--30 hours), such a fast variation effect can not be found.

The steady state analysis of the minimal model highlights three kinds
of heat stress depending on the applied temperature $T$ : normal
stresses for $T<$ 42$^\circ$C~display a pseudo thermal adaptation,
acute stresses for $42^\circ C<T<43.3^\circ C$ increase the
transcription of chaperones, and chronic stresses for $T \ge 43.3^\circ C$ induce an
accumulation of misfolded proteins over time.  Surprisingly, the
numerical values for the thresholds are highly conserved over
parameters estimations.  Moreover, a more detailed model adjusted on the
same data set, highlights also the three
regimes~\cite{sriram2012detailed}. All together,  these results  suggests a strong correlation between the existence of these three stresses regimes and the experimental data set.

Similarly, the values obtained for the concentrations to the
equilibrium temperature (37\degC) are perfectly compatible with
expectations (low values of misfolded proteins; most part of HSF1 in
complex form with HSP; significant concentration of monomeric HSP to
fold newly synthesized proteins), while no selection criterion is set
on this point. This global coherency of the 
results indicates that the proposed model is well adapted to fit the available 
experimental data set.

In single cell experiments, the kinetics of nuclear stress bodies are
monitored \textit{via} time lapse microscopy of genetically modified
cell lines to express a HSF1 protein fuse with a fluorescent tag
(\textit{e.g.} GFP). This genetic modification induces a constitutive overexpression of \textit{hsf1}, that does  not  alter significantly the dynamics, based on the HSRN results. 
Therefore, fluorescent fuse proteins constitutes  a valuable methods to investigate the heat shock response.  In contrast to this, overexpressed HSP shift the regime thresholds, so, one has to be careful in using fluorescent reporter of HSP.


At this stage, the HSRN can evolve in two ways, by the refinement in
the modeling of the heat shock response, or by the deepening of the
crosstalk with other major genetic networks. And for this second topic,
the simplicity of the present model is a clear advantage. The
refinement requests further experimental studies, for instance monitor
the responses to short-term stress (of a few minutes maximum) in order
to probe the importance of the dynamics of rapid mechanisms neglected
here (such as phosphorylation and translocation of HSF1, RNA dynamics
of HSP \dots). In the deepening of the crosstalk, one can mention (1)
the cell cycle (\textit{via} the interactions between HSF1--P53,
HSP--P21 ) to seek for the cell viability and refines the description
of thermotolerance, (2) the circadian clock (\textit{via} HSF1 induced
down regulation of $\mathrm{BMAL1}:\mathrm{CLOCK1}$ transcriptional
activity) to investigate the thermal driving of the circadian clock,
and (3) the oxidative stress response.

For instance, we know that oxidative and thermal stress responses are
tightly intricate~\cite{Ahn2003}, and the link between the two
genetics networks appears to spread across various time scales. 
At a fast time scale (within a minute), one can mention the induction by JNK (an oxidative stress response
kinase) of an hyper-phosphorylation of HSF1 that increases HSF1 transcriptional activity~\cite{JCB:JCB1163}. 
Similarly, it is known that NAD+ is the limiting fuel for the SIRT1
activity~\cite{Houtkooper2012} which  enhances the HSF1 binding on HSE~\cite{Westerheide2009}, yet  the couple (NAD+,NADH) is also 
the primary buffer of Reactive Oxygen Species (ROS) created by a oxidative
stress (upon oxidative stress the NAD+/NADH ratio increases). 
Lastly, at a longer time scales (several hours), one can mention 
the induction of a translocation of the transcriptional factor FOXO into the
nucleus (by phosphorylation by JNK)~\cite{Calnan2008}, where FOXO induces the
sirt1 transcription after behind deacetylated by SIRT1 itself~\cite{Brunet2004,Houtkooper2012}. Obviously, the time scale
of this last mechanism is much more longer than the first two, due to the transcription-translation step involved.  In the framework developed here, all
these pathways could be characterized by modification of the
regulation threshold $P_0$ to mimic the modification of HSF1
transcriptional activity.


\section*{Author Contributions} 
Designed and performed research: AS, EC, QT. Wrote the paper AS, EC, QT.

\section*{Acknowledgments} 
This work has been supported by Ministry of Higher Education and 
Research, Nord-Pas de Calais Regional Council and FEDER through the 
Contrat de Projets \'Etat-R\'egion (CPER) 2007 2013. 
Thanks to Benjamin Pfeuty for careful reading of the manuscript.  
 
\bibliography{Thommen} 
 
\clearpage

\end{document}